\documentclass{article}

\usepackage{amssymb,amsmath}
\usepackage{authblk}

\usepackage{algpseudocode,algorithm,algorithmicx}

\usepackage{graphicx}
\usepackage[T1]{fontenc}
\usepackage[utf8]{inputenc}
\usepackage{authblk}
\usepackage{tikz}
\usepackage{todonotes}
\usepackage{subfigure}

\usepackage{amsthm}

\newcommand{\R}{\mathbb{R}}
\newcommand{\CC}{\mathcal{C}}
\newcommand{\X}{\mathbf{X}}

\date{\today}

\begin{document}
	


\title{Multivariate cumulants in outlier 
	detection for financial data analysis}

\author[1]{Krzysztof Domino\thanks{kdomino@iitis.pl, ORCID 0000-0001-7386-5441}}

\affil[1]{Institute of Theoretical and Applied Informatics\\ 
	Polish Academy of Sciences\\ 
	Ba{\l}tycka 5, 44-100 Gliwice, Poland}

\maketitle

\begin{abstract}
	
There are many research papers yielding the financial data models, where 
returns are tied
either to the fundamental analysis or to the individual, often 
irrational, behaviour of investors. In the second case the bubble followed by  
the crisis 
is 
possible on the market. Such bubble or crisis is reflected by the 
cross-correlated extreme 
positive or negative returns of many assets. Such returns are modelled by the 
copula with the meaningful 
tail 
dependencies. The typical model of such cross-correlation provides the 
t-Student copula. The author demonstrates that the mutual information tied to 
this 
copula can be measured by the 4th order multivariate cumulants.
Tested on the artificial data, the 
4th order multivariate cumulant approach was used 
successfully for the financial crisis detection. For this end the author 
introduces the outliers detection algorithm. In addition this algorithm 
displays the potential application for the 
crisis prediction, where the cross-correlated extreme events may appear before 
the 
crisis in the analogy to the auto-correlated ones measured by the Hurst 
Exponent. 
\end{abstract}

\section*{Keyword}
financial crisis detection; cross-correlation of extreme returns; 
t-Student 
copula; 4th order multivariate cumulants; mutual information.


\section{Introduction}

Financial data such as prices or 
returns of assets are often non-Gaussian distributed, 
see~\cite{bouchaud2000theory}. This is in contrast with the Central Limit 
Theorem 
(CLT), which yields a Gaussian distribution, under the condition that 
increments 
are independent, stationary and with finite variance. However the 
Autoregressive Conditional Heteroskedasticity (ARCH) model of financial data 
\cite{akgiray1989conditional} violates the independent and stationary 
conditions of the CLT. In the simple ARCH model the variance of the increment 
at given time is a function of prior realisations of increments.

For a point of view of a physicist, it is interesting to look into the 
complex dynamics of a financial system and analyse the specific forms of  
its probabilistic and stochastic models. Refer to the 
review 
papers~\cite{vasconcelos2004guided, matsushita2003exponentially}, where the 
exponentially tailed 
distributions are used to 
model financial data, while the returns are expected to be long-range 
auto-correlated, especially 
near the crisis. These auto-correlations are modelled by 
the sub-diffusive 
Fractional Brownian Motion model and detected by the Hurst 
Exponent~\cite{vandewalle1997coherent, grau2000empirical}. 
Such crisis prediction scheme refers to the idea of a scale invariance of a 
financial system in the
analogy to a complex physical system~\cite{mandelbrodt2013} and log-periodic 
oscillations occurring just before the crash~\cite{vandewalle1987}.

Based on these analogies, the Hurst Exponent 
has become the meaningful tool for a 
financial crisis prediction, see~\cite{grech2008local, grech2004can}. It is 
worth to mention the author's paper, where the Hurst Exponent calculated by 
the means 
of the local Detrended Fluctuation Analysis (DFA) was used successfully to 
predict both the changes in the trends on the Warsaw Stock 
Exchange~\cite{domino2011use}, and
the global maximum that occurred on the Warsaw Stock Exchange on the $29$th 
October 2007~\cite{domino2012use}.
The recent developments of the applications 
of the Hurst Exponent in financial data analysis go toward the multi-fractality 
and the Tsallis statistics, see references in~\cite{rak2018quantitative}. 

For the `microscopic' justification of the analogy between a financial system 
and a complex physical system, refer to the `Bak Paczuski Shubik 
Model'~\cite{bak1997price} where the interplay between `rational' traders 
(performing  fundamental analysis) and `noise' traders (following others) is 
discussed by the means of a quantum field model of diffusing 
and annihilating particles. In the phase where `noise' traders dominate, the 
variation of prices would be large and a bubble and 
a crisis would occur. Based on the `Bak Paczuski Shubik 
Model', the author has developed the simple econophysical model, addressing 
investment strategies upon the random-market tension, see Section $2.2$ and 
Fig.$3$ in~\cite{kruszewska2013}. Here the tension from `noise' traders would 
cause a sub-diffusive `growth' of the market, typical to a period prior to a 
crisis. The review 
paper~\cite{sornette2009dragon} refers to large stock crashes as being
analogous to `critical points' studied by the statistical physics. These 
crashes 
are caused by a cooperative behaviour of the traders following each other 
(`noise' 
traders). 

Referring further to~\cite{sornette2009dragon}, financial data returns follow 
a distribution with power law tails. The crisis is consistent with the
auto-correlated 
extreme events and can be detected by analysing the scale invariance. During 
the 
crisis, the returns (called the `Dragon-Kings') are auto-correlated, but they 
do 
not 
necessary diverge from the power law tail of the distribution. In the current 
paper, 
the author intends to investigate the `Dragon-Kings' as the cross-correlated 
extreme 
events, i.e. the extreme drops in many assets. Here, the detection of such 
`Dragon-Kings' will concentrate on the presence of the simultaneous extreme 
events, even if the marginal distributions and the overall cross-correlation of 
marginals are unchanged. To model such scenarios we use the copula approach.

The appropriate analysis of the cross-correlations between financial assets 
is of great importance, due to the practical application in the portfolio's 
optimisation. 
Such analysis is the common domain of mathematics, finances and econophysics, 
see~\cite{de2004measuring, szego2002measures, calsaverini2009information} for 
the copula approach and 
~\cite{sornette2000nonlinear, domino2017use, rubinstein2006multi} for 
the multivariate cumulants/moments approach. 
An interesting (non-canonical) method of the analysis of the auto-correlations 
and 
the cross-correlations 
of financial data is a random matrix 
approach~\cite{sawa2015alternative}. 
However, the econophysical state of art is the Multifractal Detrended 
Cross-Correlation Analysis (MF-DCCA)~\cite{cao2018multi, he2011multi}. It 
demonstrates the power-law cross-correlations between assets, 
suggesting that the large (positive or negative) return in one asset is more 
likely to be 
followed by the large (positive or negative) return in another asset. The 
alternative representation of such observation is the copula's tail dependency. 
The 
upper tail dependency determines if the simultaneous extreme high 
events on two marginals (or more given the generalisation) are possible, the 
lower tail dependency concerns the simultaneously appearing extreme low 
events~\cite{cherubini2004copula}.

To 
demonstrate the link between 
the auto-correlations
analysis of financial data and the copula approach, 
refer to the author's papers on the Warsaw Stock Exchange~\cite{domino2014use} 
and on
major stock exchanges~\cite{domino2015use}. There, various copulas 
were used to 
model the cross-correlations of high negative returns. The choice of the copula 
was 
determined by the
auto-correlations of returns measured by 
the Hurst Exponent computed by the means of the local DFA. In the sub-diffusive 
phase, where the crisis is likely, the data 
were modelled by the copula with non-zero lower tail dependency, because 
extreme negative returns may occur simultaneously on more than one asset. 
In the normal-diffusion phase, the data were modelled by the copula with the 
zero 
lower tail dependency, as extreme negative returns may not occur simultaneously 
for 
more than one asset). Here, the `Dragon-Kings' would belong to 
the data subset modelled by the copula with the meaningful non-zero lower tail 
dependency.

Cross-correlations of financial assets can be modelled by various 
copulas~\cite{de2004measuring, szego2002measures, cherubini2004copula, 
nelsen1999introduction}. In most cases, these are: the Gaussian copula 
(derived from the Gaussian multivariate distribution), the t-Student copula 
(derived from the
t-Student multivariate distribution), one of Archimedean copulas. Among the
Archimedean 
copulas the most popular are: the Gumbel copula, the Clayton copula, and the 
Frank copula. The Gaussian copula is 
parametrised by the correlation matrix $\Sigma$, and has zero upper and lower 
tail 
dependencies. 
The t-Student copula is parametrised by the analogical $\Sigma$ matrix and the 
discrete 
parameter ($\nu$) and has non-zero $\nu$ dependent upper and lower tail 
dependencies. 
The t-Student copula is symmetric in 
the sense that for a given pair of marginals, the upper tail dependency equals 
to 
the lower one. As discussed earlier,
the crisis (the correlated extreme negative returns) is often preceded by the 
bubble 
(the correlated extreme positive returns). 
Both the Clayton and the Gumbel copula are not symmetric in this sense, and 
they 
are applicable to the analysis of the subset of high or low 
returns~\cite{domino2014use, domino2015use}. 
The Frank copula has zero both tail dependencies, analogically to the Gaussian 
one.
Archimedean copulas are parametrised by the single number, describing the 
cross-correlation 
between all marginals. Hence, to go beyond a pair-wise analysis, one 
needs either the nested Archimedean copula approach or the Vine tree 
approach~\cite{nelsen1999introduction}.
For the practical discussion refer to~\cite{semenov2017portfolio}, where 
Gaussian, t-Student and Vine copulas were compared in the multi-asset 
investment 
portfolio construction on particular data. Although the analysis was in favour 
of the Vine copula, such approach corresponds to the tree of dependencies, 
where 
the various combinations of marginals can be modelled by various copulas. As 
such, 
there are a lot of parameters (including choices of copulas). Referring to the 
discussion in this paragraph, the author 
will concentrate, in the current paper, on the use of the t-Student 
copula for the crisis period and the Gaussian copula for the
normal period.

The additional argument for the use of the t-Student copula, is its elegant 
entropic representation. In paper~\cite{calsaverini2009information}, the 
Kullback-Leibler divergence was used to measure the mutual information tied to 
the cross-correlation modelled by the t-Student copula for
financial data analysis. Given the t-Student copula, 
such mutual information (that is marginal independent) can be split into the 
first 
part tied to the copula's $\Sigma$ parameter (call it the `Gaussian' 
correlations), 
and the second part tied to the copula's $\nu$ parameter. For 
a technical computation, see 
also~\cite{opper2001advanced}. This observation gives a motivation for 
the multivariate cumulants approach. To analyse multivariate Gaussian 
distributed 
data, modelled by the Gaussian copula, one can 
use the second order multivariate cumulant (the covariance 
matrix)~\cite{duda2012pattern}. For non-Gaussian multivariate data (where 
the copula may be other than the Gaussian one), at least 
some multivariate cumulants of the order higher than $2$ are 
non-zero~\cite{mccullagh1987tensor, kendall1946advanced}. These cumulants, 
called the higher-order multivariate cumulants, can carry the meaningful 
information about non-Gaussian distribution or non-Gaussian copula. 
The higher-order 
multivariate cumulants can 
be represented in the form 
of the $d$-mode arrays called the \emph{higher-order cumulans 
tensors}. For further discussion on such higher-order cumulants tensors 
see~\cite{domino2017tensorsnet} and references within. 

The higher-order multivariate cumulants approach is somewhat similar to 
the higher-order multivariate moments approach, used in the investment 
portfolio's optimisation~\cite{rubinstein2006multi, arismendi2014monte, 
	jondeau2015moment}. However, the higher-order multivariate moments are 
	non-zero 
	for Gaussian 
multivariate data and their application for the analysis of information tied to 
the non-Gaussian distribution (or copula) is not straight-forward. 
For an econophysical application of the higher-order multivariate cumulants in 
investment portfolio's optimisation refer 
to~\cite{sornette2000nonlinear}. Here the method of the even order 
cumulants determination, given some probabilistic models (referring to a 
Weibull 
distribution), was developed using an analogy to the Feynman diagrams.

Observe that the multivariate cumulants are both copula's dependnet and 
univariate 
marginal distributions dependent, hence the aspect second must be discussed as 
well.
Financial data often have non-Gaussian univariate marginal 
distributions. Most common are: the t-Student~\cite{de2004measuring, 
eckhard2008} one, the 
Johnson one~\cite{cayton2015time}, the Log-normal~\cite{rachev2003handbook} ane 
and the Extreme Value Distribution~\cite{domino2014use}. Recall as well the 
power law 
tails of distributions mentioned 
in~\cite{sornette2009dragon} that may be followed both during a crisis and 
non-crisis times. This is an argument for having unchanged univariate 
distributions 
and focusing on various copulas given various market's stages.
There are two important arguments for using the t-Student marginal 
distributions: first, this distribution was widely used to model financial data 
(see 
for example~\cite{de2004measuring}); second, 
it has the power-law tails and is symmetric (the extreme drops will be modelled 
similar to the extreme increases).

Following~\cite{rubinstein2006multi, jondeau2015moment}, in 
paper~\cite{domino2017use}, the author has used the Higher Order Singular Value 
Decomposition  (HOSVD) of cumulants tensors for portfolio's optimisation. 
Here the use of cumulants of order $2,3, \ldots, 6$ computed for daily 
percentage returns of shares traded on the Warsaw Stock Exchange gave safer 
portfolios for the crisis time. This crisis time was predicted by means of 
the 
Hurst Exponent computed for the stock exchange index by means of the local DFA.
For technical issues concerning the
HOSVD refer to~\cite{de2000multilinear}, while for the detailed algorithms for 
computing the higher-order cumulants tensors see~\cite{domino2017tensorsnet} 
and~\cite{domino2017updates} in the case of streams of (financial) data.

The working hypothesis will be: can one detect crisis events 
and predict the crisis by the use of the higher-order multivariate 
cumulants calculated for 
financial data. Such crisis detection and crisis prediction will not analyse 
the auto-correlations of financial data and can be complementary to priorly 
mentioned methods that analyse the auto-correlations. The author uses the 
outliers 
detection scenario by means of the modified (simplified) algorithm 
form~\cite{pena2001multivariate}. Such approach concentrates on 
the projections of data into directions 
with high values of the higher-order cumulant of the given order.

The algorithms are implemented in the \texttt{Julia} programming language 
\cite{bezanson2014julia} that is efficient, open source and high level 
programming language suitable for scientific computations. Implementation of 
these algorithms is available on the GutHub repository \cite{cumf}. For 
implementation of an algorithm for artificial data generation see
\cite{cop}.

The paper is organised as follows. In Section~\ref{s::prel} the 
probabilistic model of data is discussed, in particular in 
Subsection~\ref{s::elcops} t-Student and Gaussian 
copulas are introduced, in Subsection~\ref{s::mcum} multivariate cumulants of 
such models are discussed, while in Subsection~\ref{s::hosvd} these cumulants 
are used to project data on particular directions.
In Section~\ref{sec::outdet} the outliers detection is discussed, in 
Subsection~\ref{sec::rx} the state of art Reed-Xiaoli Detector is discussed, in 
Subsection~\ref{sec::c4det} the cumulants based detector introduced by an 
authors is discussed, while in Subsection~\ref{sec::da} these detectors are 
tested on artificial data. In Section~\ref{sub::fint} real live financial data 
are analysed in crisis detection and crisis prediction scenarios.

\section{Data model}\label{s::prel}

In this section, the probabilistic model 
of data, including the copula, is discussed. Further, 
the higher-order multivariate cumulants are introduced, as the tools applicable 
to collect information about the cross-correlations modelled by the copulas.

\subsection{Copulas}\label{s::elcops}

By the Sklar theorem~\cite{sklar1959}, each multivariate Cumulative Density 
Function (CDF) 
$\mathbf{F}$ can be split onto 
the univariate CDFs and the copula $C$. The copula describes 
the particular cross-correlation between the marginals.
For the introduction of the probabilistic model of data, let $\mathbf{v} = 
[v_1, 
\ldots, v_n] 
\in \R^n$ be a 
single realisation of the $n$-variate real valued random vector, and let $F_i$ 
be the
$i$th continuous univariate marginal CDFs. The multivariate (joint) CDF 
is~\cite{nelsen1999introduction}:
\begin{equation}\label{eq::co}
\mathbf{F}(\mathbf{v}) = C\left(F_1(v_1), \ldots, F_n(v_n)\right).
\end{equation}
Differentiating the multivariate CDF, the multivariate Probability Density 
Function (PDF) is:
\begin{equation}\label{eq::cod}
\mathbf{f}(\mathbf{v}) = c\left(F_1(v_1), \ldots, F_n(v_n)\right) 
\prod_{i = 1}^n f_i(v_i),
\end{equation}
where
\begin{equation}
c(\mathbf{u}) = 
\frac{\partial^n}{\partial u_1 \ldots \partial 
	u_n} C(\mathbf{u})
\end{equation} 
is the copula density, and $f_i(v_i) = \frac{d}{d v_i} F_i(v_i)$ are 
the univariate PDFs.
Let $\mathbf{u} = [u_1, \ldots, u_n] \in [0,1]^n$ be a 
single realisation of the $n$-variate random vector with uniform marginals on 
$[0,1]$, such that $u_i = F_i(v_i)$. In the probabilistic terms, the copula 
$C(\mathbf{u}): [0,1]^n \rightarrow [0,1]$ is the (joint) multivariate CDF 
with all uniform 
marginals on $[0,1]$. 
Hence the copula~\cite{nelsen1999introduction} is 
marginal independent and defines the cross-correlation between the marginals. 

For analysing cross-correlated extreme events, one can refer to the tail 
dependencies~\cite{nelsen1999introduction}. The tail 
dependency determines if the simultaneous extreme 
events on two marginals (or more in a generalised case) are 
possible. Given the bivariate copula $C(u_1, 
u_2)$, the lower ($\lambda_l$) and the upper ($\lambda_u$) tail dependencies 
are:
\begin{equation}
\lambda_l = P(u_1 \rightarrow 0 |u_2 \rightarrow 0) \text{ and } \lambda_u = 
P(u_1 
\rightarrow 1 |u_2 \rightarrow 1).
\end{equation} 

The t-Student and the Gaussian copulas are both derived from the t-Student 
$\mathbf{T}_{\nu, \Sigma}(\mathbf{v})$ and the Gaussian  
$\mathbf{G}_{\Sigma}(\mathbf{v})$ multivariate CDFs, that 
are~\cite{kotz2004multivariate} (assuming zero mean): 
\begin{equation}\label{eq::tcdf}
\mathbf{T}_{\nu, \Sigma}(\mathbf{v}) = \int_{-\infty}^{v_1} \ldots 
\int_{-\infty}^{v_n} \mathbf{t}_{\nu, \Sigma}(y_1, \ldots, y_n) dy_1 \ldots 
dy_n 
\end{equation}
where
\begin{equation}
\mathbf{t}_{\nu, \Sigma}(\mathbf{v}) = \frac{\Gamma(\frac{\nu + 
		n}{2})}{\Gamma(\frac{\nu}{2})\nu^{n/2} \pi 
	^{n/2}|\Sigma|^{1/2}}\left(1+\frac{\mathbf{v}
	\Sigma^{-1}\mathbf{v}^{\intercal}}{\nu}\right)^{-\frac{\nu+n}{2}},
\end{equation}
and
\begin{equation}\label{eq::gcdf}
\mathbf{G}_{\Sigma}(\mathbf{v}) = \int_{-\infty}^{v_1} \ldots 
\int_{-\infty}^{v_n} \mathbf{g}_{\Sigma}(y_1, \ldots, y_n) dy_1 \ldots dy_n
\text{, }
\mathbf{g}_{\Sigma}(\mathbf{v})= \frac{1}{\sqrt{(2
		\pi)^n|\Sigma|}}\exp\left(-\frac{\mathbf{v}
	\Sigma^{-1}\mathbf{v}^{\intercal}}{2}\right).
\end{equation}
In both cases, the parameter $\Sigma$ is the semi-positively defined symmetric 
matrix. The t-Student copula case is also parametrised by the integer 
positive parameter $\nu$ (number of degrees of freedom). The covariance matrix 
of the Gaussian multivariate distribution equals to 
$\Sigma$, while the covariance matrix of the t-Student multivariate 
distribution equals to 
$\frac{\nu}{\nu-2} \Sigma$ for $\nu > 2$ \cite{kotz2004multivariate}.

For the 
copulas definition, it is 
convenient to assume that the parameter $\Sigma$ is a symmetric positively 
semi-defined matrix with ones on the diagonal. This assumption gives the 
standard univariate 
Gaussian CDFs, denoted by $G$, and univariate t-Student CDFs  
with variance $\frac{\nu}{\nu-2}$, denoted by $T_{\nu}$.
The t-Student copula is given by:
\begin{equation}
C_{\nu_c, \Sigma}(u_1, \ldots, u_n) = \mathbf{T}_{\nu_c, 
\Sigma}(T_{\nu_c}^{-1}(u_1), 
\ldots, T_{\nu_c}^{-1}(u_n)),
\end{equation}
where the $c$ subscript in $\nu_c$ is used to distinguish the copula parameter. 
The Gaussian copula 
is given by:
\begin{equation}
	C_{\Sigma}(u_1, \ldots, u_n) = \mathbf{G}_{\Sigma}(G^{-1}(u_1), \ldots, 
	G^{-1}(u_n)).
\end{equation}
Bivariate tail dependencies between marginals 
$i_1$ and $i_2$ for the t-Student copula are~\cite{nelsen1999introduction}:
\begin{equation}
\lambda_l = \lambda_u = 2 t_{\nu_c +1} \left( -\sqrt{\nu_c+1} 
\sqrt{\frac{1-\sigma_{i_1,i_2}}{1+\sigma_{i_1,i_2}}}\right),
\end{equation}
where $\sigma_{i_1,i_2}$ is the corresponding element of $\Sigma$. The 
univariate t-Student CDF is:
\begin{equation}\label{eq::univt}
t_{\nu_u}(x) = \frac{\Gamma(\frac{\nu_u + 
		1}{2})}{\Gamma(\frac{\nu_u}{2}) \sqrt{\nu_u 
		\pi}}\left(1+\frac{x^2}{\nu_u}\right)^{-\frac{\nu_u+1}{2}},
\end{equation}
here the subscript $u$ in $\nu_u$ is used to distinguish univariate marginal's 
parameter. Observe 
that in the 
case of $\nu_c \rightarrow \infty$ the t-Student copula becomes the Gaussian 
one. 
It is easy to see that given 
$\sigma_{i_1,i_2} < 1$, the following holds $\lambda_l = \lambda_u = 0$ for the 
Gaussian copula. Hence the model that switches between the Gaussian and the 
t-Student copulas with 
given $\nu_c$ can be used for financial data where either simultaneous extreme 
events are 
possible (a crisis) or not (a normal trading period).

Finally, following Eq.~\eqref{eq::cod} given the t-Student copula and the 
t-Student 
univariate marginals, we have the t-Student multivariate PDF only if $\nu_c = 
\nu_u$ for all marginals.
%
%

Copulas determine the relationship between marginals. To investigate 
the information of such relationship one can use the relative entropy, called 
also the Kullback-Leibler divergence or the mutual 
divergence~\cite{kullback1951information}. 
The relative entropy is originated from the variational `Mean Field' 
approach~\cite{opper2001advanced}, where it measures a divergence of the model 
from the basic distribution. 

In~\cite{calsaverini2009information} the multivariate PDFs with the 
Gaussian copula or the t-Student copula 
$C = (C_{\Sigma}, C_{\nu_c, \Sigma})$ are analysed. 
The basic distribution has the 
same marginals and the independent (product) copula $C_{\perp}(\mathbf{u}) = 
\prod_i u_i$, yielding the following Kullback-Leibler divergence:
\begin{equation}
I(C, C_{\perp}) = \int d\mathbf{v} \ \mathbf{f(v)} \log \left( 
\frac{\mathbf{f(v)}}{\prod_{i} f_i(v_i)} \right).
\end{equation}
Such divergence is supposed to measure information stored in the relationship 
between marginals, and as such 
it is marginal independent. 

In~\cite{calsaverini2009information} it was shown that given the t-Student 
multivariate distribution (hence the t-Student copula) the mutual information 
can be split onto the `Gaussian part' $I_{\Sigma} = -\frac{1}{2} \log 
\left(\det (\Sigma) \right)$ and the $\nu_c$ - dependent part:
\begin{equation}\label{eq::Inun}
I_{\nu_c, n} = \log \left( \frac{\left(\beta\left(\frac{\nu_c}{2}, 
\frac{1}{2}\right)\right)^n \Gamma\left(\frac{n}{2}\right)}{\pi^{\frac{n}{2}} 
\beta\left(\frac{\nu_c}{2}, 
\frac{n}{2}\right)} 
\right) - \frac{\nu_c (n-1)}{2} \psi\left( \frac{\nu_c}{2} \right) +  \frac{n 
(\nu_c 
+ 1)}{2} \psi\left( \frac{\nu_c+1}{2} \right) - \frac{\nu_c 
+ n}{2} \psi\left( \frac{\nu_c+n}{2} \right),
\end{equation}
that measures the additional information due to the fact that the 
copula is the t-Student one rather than the Gaussian one. Here $\psi$ is the 
digamma function and 
the whole Kullback-Leibler divergence is:
\begin{equation}
I(C_{\nu_c, \Sigma}, C_{\perp}) = I_{\Sigma} + I_{\nu_c, n}.
\end{equation}

Observe that $I_{\nu_c, n} = 0$ for $\nu_c \rightarrow \infty$, i.e. for the 
Gaussian copula. Hence, the author is going to investigate whether $I_{\nu_c, 
n}$ can be detected by the means of higher-order multivariate cumulants. On the 
one hand, these 
cumulants can be efficiently computed~\cite{domino2017tensorsnet, 
domino2017updates}, and on the other hand they are applicable in 
financial data analysis~\cite{domino2017use, martin2013consumption}.

Referring to Eq.~\eqref{eq::cod} given the copula (t-Student or Gaussian) 
univariate  marginal distributions need to be selected as well, to have a full 
probabilistic model. In 
papers~\cite{de2004measuring, 
	eckhard2008} the t-Student univariate PDFs were used to model financial 
	data, and especially their log increments. The parameter $\nu_u$ was in 
	range $[4,6]$ in most cases of the well-established markets. Hence, for 
	sake of 
	the analysis, the $\nu_u = 6$ will be selected for all marginals, although 
	the generalisation 
	for other values of $\nu_u$ is straightforward.
It is easy to observe that univariate t-Student distribution has 
	the asymptotic power law tails, which is consistent with the financial data 
	model in~\cite{sornette2009dragon}. For $\nu_u \rightarrow \infty$ it 
	becomes 
	the Gaussian univariate distribution.

\subsection{Higher-order multivatiate cumulants}\label{s::mcum}

The $d$\textsuperscript{th}-order multivariate cumulant of marginals 
denoted by the multi-index $\mathbf{i} = (i_1, 
\ldots, i_d)$ is defined by the following generator function:
\begin{equation}
\R \ni c_{i_1, \ldots, i_d} = (-\text{i})^d \frac{\partial^d}{\partial 
y_{i_1}, \ldots, \partial y_{i_d}} \log \phi(\mathbf{y}) \big|_{\mathbf{y} = 0}
\text{ where } \phi(\mathbf{y}) = \int_{\R^n} e^{\text{i} 
\mathbf{y} \cdot \mathbf{v}^{\intercal}} \mathbf{f}(\mathbf{v}) d \mathbf{v},
\end{equation}
and $\mathbf{f}(\mathbf{v})$ is the multivariate PDF. To compute such cumulants 
it is 
convenient to compute multivariate moments 
first and then use Mc Cullagh relation~\cite{mccullagh2009cumulants} or its 
simplification~\cite{domino2017tensorsnet}.
The $d$\textsuperscript{th} multivariate moment is \cite{kendall1946advanced}:
	\begin{equation}\label{eq::mom}
	m_{i_1, \ldots, i_d} = \int_{\R^n} v_{i_1} \cdot \ldots \cdot v_{i_d} 
	\mathbf{f}(\mathbf{v}) d 
	\mathbf{v}.
	\end{equation}
If all univariate marginals are zero mean, the moment becomes the central 
 moment denoted by: $\tilde{m}_{i_1, \ldots, i_d}$.

There are several interesting observations. Given $n$ marginals the $d$-order 
moment or the 
$d$-order cumulant can be organised in the form of tensor (the $d$ mode array) 
of size $\R^{n 
\times \ldots \times n}$. Each moment's or cumulant tensor is symmetric in 
the sense, that the order of indexing $i_1, \ldots, i_d$ of its elements does 
not matter. For 
the cumulant tensor of order $d$ the notation $\CC_d$ will be used. 
Following~\cite{mccullagh2009cumulants} multivariate cumulants of order $2-4$ 
are given by:

\begin{itemize}
	\item $c_{i_1, i_2} = \tilde{m}_{i_1, i_2}$,
	\item $c_{i_1, i_2, i_3} = \tilde{m}_{i_1, i_2, i_3}$,
	\item $
		c_{i_1, \ldots, i_4} = \tilde{m}_{i_1, \ldots, i_4} - 
		\tilde{m}_{i_1, i_2}\tilde{m}_{i_3, i_4} - \tilde{m}_{i_1, 
		i_3}\tilde{m}_{i_2, i_4} - \tilde{m}_{i_1, 
			i_4}\tilde{m}_{i_2, i_3}$.
	\end{itemize}

\begin{figure}
	\subfigure[$\nu_c = 6$, various cross-correlation $\sigma$.
	\label{fig::tcoprho}]{\includegraphics[width=0.32\textwidth]{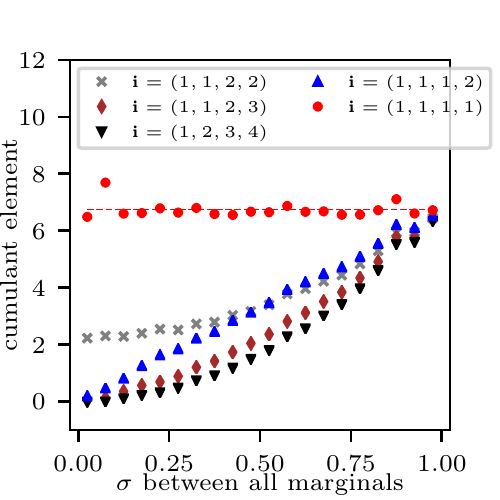}}
	\subfigure[Various $\nu_c$, constant $\sigma = 0.5$ 
	between all marginals. 
	\label{fig::tcopnu}]{\includegraphics[width=0.32\textwidth]{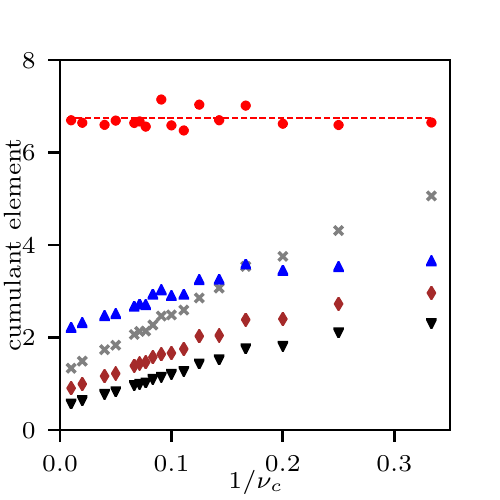}}
		\subfigure[The addtional mutual information $I_{\nu_c, 
		n}$, see 
		Equation~\eqref{eq::Inun}.\label{fig::tentropy}]{\includegraphics[width=0.32\textwidth]{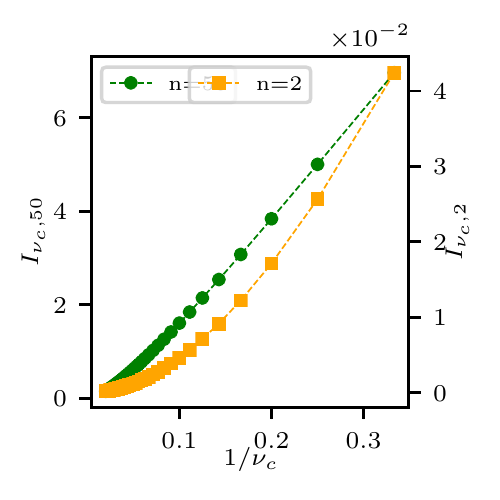}}
	\caption{Elements of the $4$th cumulant tensor and 
	mutual information for the
	t-Student 
	copula parametrised by $\nu_c$ and $\sigma$ -the parameter of the 
	cross-correlation 
	between all marginals. t-Student univariate marginals with $\nu_u=6$ 
	parameter were used for all marginals. The case $\frac{1}{\nu_c} 
	\rightarrow 0$ refers to the Gaussian copula.}
	\label{fig::tcop}
\end{figure}

Each higher-order multivariate cumulant tensor has elements lying on its 
super-diagonal, i.e. with indexing $i_1 = i_2 = \ldots = i_d$. These 
elements refer to the $d$th cumulant of the $i$th univariate marginal 
distribution.
Other elements are off-diagonal, these refer to multivariate cumulants 
tied both to some univariate marginals and to the copula. For particular values 
of the elements of the $4$th order cumulant tensor for the t-Student marginals 
(with $\nu_u = 6$) 
and the t-Student copula see 
Figures~\ref{fig::tcoprho}, and~\ref{fig::tcopnu}.


As the t-Student copula is based on the t-Student multivariate distribution, 
let us discuss briefly the cumulants of the t-Student multivariate 
distribution, 
see Eq.~\eqref{eq::tcdf}. 
The 
second cumulant matrix is~\cite{kotz2004multivariate}:
\begin{equation}
\CC_2 = \frac{\nu}{\nu-2} \Sigma
\end{equation}
For the t-Student multivariate distribution, the multivariate cumulant of order 
$3$ 
is zero \cite{ kotz2004multivariate, balakrishnan1998note}. This is caused by 
certain sort of the symmetry of the t-Student distribution. 
Hence, for  
the t-Student copula with symmetric univariate marginals $\CC_3 = 0$ is 
expected as well. 

Although diagonal (univariate) elements of the $4$th cumulant of t-Student 
distribution are $\frac{6}{\nu-4} \left(\frac{\nu}{\nu-2}\right)^2$, 
off-diagonal elements are more complex, but positive, see 
Figures~\ref{fig::tcoprho}, and~\ref{fig::tcopnu}.

For the Gaussian copula ($\nu_c \rightarrow \infty$), we expect the $4$th order 
multivariate cumulant to be marginal dependent only, see 
Figure~\ref{fig::tcoprho}. Note the linear 
relation between the $4$\textsuperscript{th} cumulant tensor elements and 
$\frac{1}{\nu_c}$ in 
Figure~\ref{fig::tcopnu}, 
for theoretical justification one should refer to a impact of this parameter of 
the t-Student multivariate 
distribution on the $4$th multivariate cumulant  
\cite{kotz2004multivariate}. 

Compare Figure~\ref{fig::tcopnu} with 
additional mutual information $I_{\nu_c, n}$ (see 
Equation~\eqref{eq::Inun}) of the t-Student copula in 
Figure~\ref{fig::tentropy}, here for large number of marginals - $n$ the linear 
relation appears as well. Concluding, the elements of $\CC_4$ seems to be the 
good 
features of the additional mutual information of the t-Student copula 
(parametrised by 
$\nu_c$) with t-Student 
marginals (parametrised by $\nu_u$).
Although $I_{\nu_c, n}$ is $\sigma$ independent, the $4$th order cumulants 
elements 
are (parabolically) increasing with $\sigma$, see Figure~\ref{fig::tcoprho}. 
Hence, 
in the outlier selection, see Algorithm~\ref{alg::det}, some normalisation is 
desirable. The 
normalisation of the decision parameter by the Median Absolute Deviation (MAD) 
is 
preformed in the line $12$ of the algorithm, while in the line $10$ of the 
algorithm the Kurtosis (the $4$th univariate cumulant, normalised by the square 
of the variance) is used in the stop condition.

\subsection{The analytical tools}\label{s::hosvd}

For the practical reason of the outliers detection for the financial data, the 
projection of data onto directions with high values of the given higher-order 
cumulant is used. For this reason, the Higher Order 
Singular Value Decomposition (HOSVD)~\cite{de2000multilinear} of the higher 
order cumulants tensor can be used, see~\cite{rubinstein2006multi} for 
moment approach on financial data. For the review of other methods of tensor 
decomposition (including algorithmic implementation) 
see~\cite{schatz2014exploiting, 
kolda2009tensor}.

Following~\cite{de2000multilinear, tucker1966some} and the discussed symmetry 
of the cumulant tensor (multi-index permutation invariance) such 
decomposition can be performed by the eigenvalue decomposition of the 
particular matrix. Let 
$c$ be the element of $\CC_d$, its contraction with itself in $d-1$ modes gives 
the symmetric, (semi) positively defined matrix $\mathbf{M^{(d)}} \in \R^{n 
\times 
n}$,

\begin{equation}\label{eq::md}
m^{(d)}_{j_1,j_2} = \sum_{i_2, \ldots, i_d = 1,\ldots, 1}^{n, \ldots, n} 
c_{j_1, i_2, 
\ldots, i_d} \cdot 
c_{j_2, i_2, \ldots, i_d}.
\end{equation}
Due to the mentioned symmetry, it is not important in which modes the 
contraction 
is performed, hence the matrix $\mathbf{M^{(d)}}$ is uniquely defined.
By laborious element-wise operations one can show that the 
eigenvalues/eigenvectors decomposition of $\mathbf{M^{(d)}} = \mathbf{W} 
\mathbf{D} \mathbf{W}^{\intercal}$ corresponds to the HOSVD of the original 
 tensor $\CC_d$, given it is symmetric, here the matrix $\mathbf{D}$ is the 
 diagonal matrix of non-negative eigenvalues 
of $\mathbf{M^{(d)}}$. 

Further, following~\cite{de2000multilinear}, it can be shown that if the 
eigenvalues 
in $\mathbf{D}$ are ordered from highest to lowest, the first columns of 
$\mathbf{W}$ (that are orthonormal) project data on the directions with high 
absolute value of the $d$-order cumulant. By analogy, the last columns of 
$\mathbf{W}$ 
project data onto the directions with low absolute value of the $d$-order 
cumulant.
The second observation was used in~\cite{domino2017use}, for the determination 
of 
portfolios of financial data with low variability for the crisis. 
In~\cite{domino2017use}, the 
modification was made to construct the projection of data on the directions 
where the
cumulants of subsequent order have all low absolute values to minimise the risk.

However, in the current approach, the analysis concentrates on the directions 
with 
absolute value of the $4$th order cumulant, to detect the crisis data that are 
supposed to be modelled by the t-Student copula.

\section{The outliers detection}\label{sec::outdet}

Let us represent $t$ realisations of $n$ variate data in the following
matrix form: $\X \in \R^{t \times n}$.
The subset of outliers realisations of size $\tau$ is approximately an order of 
magnitude smaller than the size of data ($t$). In this scenario, the outliers 
are 
expected to 
have the different probabilistic model than the ordinary data. For the outliers 
detection, the author will use the $4$th order multivariate cumulant outlier 
detector 
(the modified detector form~\cite{pena2001multivariate}), and
the state of art Reed-Xiaoli (RX) 
detector~\cite{reed1990adaptive, chang2002anomaly}.
For the current review of outlier detectors see~\cite{yang2019}. here, it was 
pointed 
out that out of automatic (unsupervised) detectors the standard 
deviation based ones, such as the RX detector, are most frequently used. 
Moreover, 
the author's intention is to demonstrate the information tied to the 
non-Gaussian joint distribution hence, the comparison with the RX detector, 
based on the mean vector and the covariance matrix, is straightforward.

\subsection{The Reed-Xiaoli Detector}\label{sec::rx}

In theory, the RX detector requires 
the ordinary data (sometimes called the background) to follow the Gaussian 
multivariate 
distribution with the fixed covariance matrix and mean vector. Outliers 
are supposed to be
modelled by the Gaussian distribution with the same covariance, but different 
mean vector. Nevertheless, the RX detector is applicable to 
analyse non-Gaussian distributed data as well, see for example  
\cite{glomb2018application} and references therein. 

To 
introduce the RX 
detector, suppose the 
$j$\textsuperscript{th} realisation of data is represented in the vector form 
$\mathbf{x}_j \in \R^n$. Having estimated the 
mean vector $\mu$ and the covariance matrix $\CC_2$ 
from data, we 
can calculate the Mahalanobis distance~\cite{chandra1936generalised} between the
given realisation and the mean 
vector:
\begin{equation}\label{eq::mdist}
	md_j = (\mathbf{x}_j - \mu) \CC_2^{-1} (\mathbf{x}_j - 
	\mu)^{\intercal}.
\end{equation}
The higher the value of $md_j$, the higher probability that the 
$j$\textsuperscript{th} realisation is an outlier.

Given the mentioned Gaussian assumptions, one can use 
the 
percentile of the 
$\chi^2$ squared distribution with $n$ degrees of freedom for the threshold 
value, to decide whether
the 
analysed piece of data is an outlier or not. 
This representation is not straightforward if data are non-Gaussian 
distributed. However, here one can still use the threshold parameter, but 
without the `elegant' probabilistic interpretation. 
Please note that, that the 
Mahalanobis distance appears in practice
as a part of many detectors of outliers that are not necessarily Gaussian 
distributed~\cite{das1986detection, 
sinha1984detection}.

\subsection{The $4$th order multivariate cumulant outlier 
detector}\label{sec::c4det}

As the higher order multivariate cumulants are supposed to carry information 
about 
outliers, these cumulants may be used for an outlier detection. For this end 
the 
modification of the outlier detection algorithm 
from~\cite{pena2001multivariate} is proposed. The original procedure can be 
summarised in the
following steps.
\begin{enumerate}
	\item Remove the mean vector and the standard cross-correlation from data 
	$\mathbf{X} \rightarrow (\mathbf{X} - \mu) \CC_2^{-\frac{1}{2}}$.
	\item Project data onto directions with the highest and the lowest $4$th 
	order 
	moment.
	\item Compute the distance between each realisation of the projected data 
	and 
	its median.
	\item If the distance is higher than the threshold, remove the sample 
	from data and move it to the outliers candidates set.
	\item Continue the procedure as long as the stop condition is fulfilled, 
	i.e. no 
	more outliers 
	candidates appear or the size of the set of outliers 
	candidates is roughly a half of the data size.
	\item Use the RX detector with the parameters estimated from the data 
	subset 
	without some outliers candidates, to detect true outliers. The detection 
	threshold is the $99$th 
	percentile of the corresponding $\chi^2$ distribution.
\end{enumerate}
In \cite{pena2007combining} 
the procedure was modified by adding some additional 
random directions to the specific directions.

The author's modification of the original outliers detection algorithm 
in~\cite{pena2001multivariate}, is 
presented in Algorithm~\ref{alg::det}. The modification concerns 
basically the replacement of the 
$4\textsuperscript{th}$ order moment by the $4\textsuperscript{th}$ order 
cumulant as being more informative for the financial data model discussed in 
this paper, see Subsection~\ref{s::mcum}. Hence, the author proposes to project 
data onto the directions with the highest 
absolute value of the $4\textsuperscript{th}$ order cumulant, see 
Subsection~\ref{s::hosvd}. Observe that low 
absolute value of the $4$th cumulant is supposed to reefer to the Gaussian 
multivariate model, far from the model of the financial crisis.  

Further, the author skips the RX detection in the last step of the 
outliers detection and the outliers candidate intermediate step. This is 
because 
both probabilistic models of data and outliers are non-Gaussian, hence the RX 
threshold parameter can not be simply estimated from the 
$\chi^2$ distribution. Further, the detector becomes simpler and is 
parametrised 
by only one parameter.

The last modification concerns the stop condition given no intermediate step of 
the outliers candidates. The stop condition must ensure that the small subset 
of data can be detected as outliers. For this sake, the author proposes the 
mean 
squared Kurtosis over specific directions. The iterative removal of 
outliers is being performed as long as such Kurtosis of remaining ordinary data 
is dropping. 

\begin{algorithm}[t]
	\caption{the $4$th order multivariate cumulant outliers detection.
		\label{alg::det}}
	\begin{algorithmic}[1]	
		\State \textbf{Input}: $\X \in \R^{t\times n}$ - data, $\beta$ - 
		the sensitivity parameter, $r$ - the number of 
		specifics directions. 	
		\State \textbf{Output:} outliers $\subset (1: t)$.
		\Function{hosvdc4detect}{$\X$, $\beta$, $r$}
		\State $\X = \X -  \mu$ \Comment{remove the mean vector} 
		\State $\X = \X \CC_2^{-\frac{1}{2}}$ \Comment{remove 
		standard cross-correlations, $\CC_2$ - the 
		covariance matrix}
		\While{$k_{\text{this step}} < k_{\text{previous step}}$} 
		\State compute $\CC_4(\X)$  and $\mathbf{M^{(4)}}$ \Comment{see 
		Eq.~\eqref{eq::md}}
		\State $W_1, \ldots, W_r =\text{eigenvec}\left(\mathbf{M^{(4)}}\right): 
		\lambda_1 \geq \ldots \geq \lambda_n$ \Comment{$W_i \in \R^n$}
		\State $Z_1 = \left(W_1^{\intercal}\right) \X, \ldots, Z_r = 
		\left(W_r^{\intercal}\right) \X$
		\State $k = \sqrt{\sum_{i=1}^r \left(\text{kurtosis}(Z_i)\right)^2}$
		\For {$j \gets 1 \textrm{ to (number of samples left)}$}
		\If{$\max_{i} \frac{|z_{j,i} - \text{median}(Z_i)|}{\text{MAD}(Z_i)} > 
		\beta$} \Comment{$z_{j,i}$ is  $j$\textsuperscript{th} element of 
		$Z_i$, MAD is the Median Absolute Deviation}
		\State append $j$ to outliers
		\EndIf
		\EndFor
		\State Remove outliers realisations from $\X$
		\EndWhile
		\State\Return outliers
		\EndFunction 
	\end{algorithmic}
\end{algorithm}

\subsection{Experiments on artificial data}\label{sec::da}

The artificial data are designed to mimic, at least to some extend, the 
financial data. Hence following~\cite{sornette2009dragon} the univariate 
marginals with the asymptotic power law tails are selected. For the particular 
analysis the t-Student univariate marginals, see Eq.~\eqref{eq::univt}, are 
selected. As mentioned before the parameter of  the marginals is
$\nu_u = 6$. Data size is proposed to be $t = 1000$ corresponding to a few 
trading years or the size of the longest observation window 
in~\cite{domino2011use}. The number of marginals is set to $n = 30$, which 
corresponds to the size of the typical market index. Data are divided onto the 
outliers subset of size $\tau = 0.1 t$ and the ordinary data subset of size $t 
- \tau$. One can expect the crisis periods to be of an order of magnitude 
shorter 
than the normal periods. Following, the discussion in the Introduction and 
Section~\ref{s::prel} the ordinary data are expected to have zero tail 
dependencies (simultaneous extreme events are not possible). On the contrary, 
the 
crisis data are expected to have the meaningfully positive tail 
dependencies between some marginals (there may be still the subset of safe 
assets). Hence, for modelling the ordinary data, the Gaussian copula is 
selected 
while for modelling the crisis data the t-Student copula for the subset 
containing one half of 
the marginals and the Gaussian copula for the rest of the marginals. 
The 
correlation matrix is randomly generated and is expected to be similar for 
ordinary data and the outliers.


The artificial data generation scheme  
uses the modified t-Student distributed data generator 
\cite{kotz2004multivariate}. The algorithm is implemented in the \texttt{Julia} 
programming 
language \cite{bezanson2014julia} and available on the GitHub repository 
\cite{cop} as the function \texttt{gcop2tstudent()}. It is summarised in the 
following points.
\begin{enumerate}
	\item Randomly generate the correlation matrix $\Sigma$, by means of the 
	'random' method 
	from~\cite{domino2018hiding} - function \texttt{cormatgen\_rand()} 
	in~\cite{cop}.
	\item Generate $t$ realisations of 
	$n$-variate Gaussian distribution parametrised by $\Sigma$.
	\item For the randomly selected subset of marginals of size $n/2$, 
	transform 
	multivariate Gaussian distribution into the t-Student one (parametrised by 
	$\nu_c$) by multiplying all elements of each 
	realisation by $\sqrt{\frac{\nu_c}{v_0}}$, where $v_0 \sim 
	\chi^2_{\nu_c}$.
	\item Transform all univariate marginals to the desirable ones, i.e.  
	the t-Student 
	with the parameter $\nu_u$.
\end{enumerate}

\begin{figure}
 	\begin{center}
 		 	\includegraphics{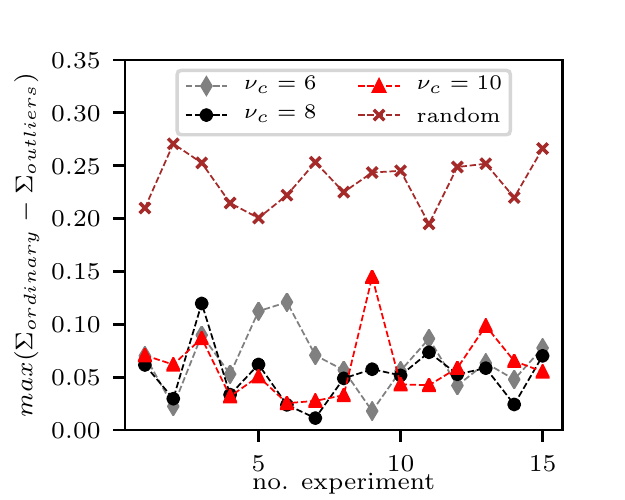}
 		\caption{Maximal differences between the correlation matrix of the 
 		ordinary data
 			$\Sigma_{\text{ordinary}}$ and the 
 			outliers $\Sigma_{\text{outliers}}$, in both cases $100$ 
 			samples were taken. Random means twice sample the $\Sigma$ matrices 
 			by means of the 'random' method from~\cite{domino2018hiding} 
 			equivalently the 
 			function 
 			\texttt{cormatgen\_rand()}~\cite{cop}.}
 		\label{fig::diff}
 	\end{center}
 \end{figure}

Although, the sampling from $\chi^2_{\nu_c}$ slightly decreases 
the cross-correlation between two subsets of marginals (see point $3$), the 
effect 
decreases with the rise of $\nu_c$, and correlation matrices of  the ordinary 
data 
($\Sigma_{\text{ordinary}}$) and the outliers ($\Sigma_{\text{outliers}}$) are 
similar, the maximal difference rarely exceeds $0.1$, see 
Figure~\ref{fig::diff}.

%

\begin{figure}
	\centering
	\subfigure[$\nu_c = 
	6$\label{fig::nu6}]{\includegraphics[width=0.32\textwidth]{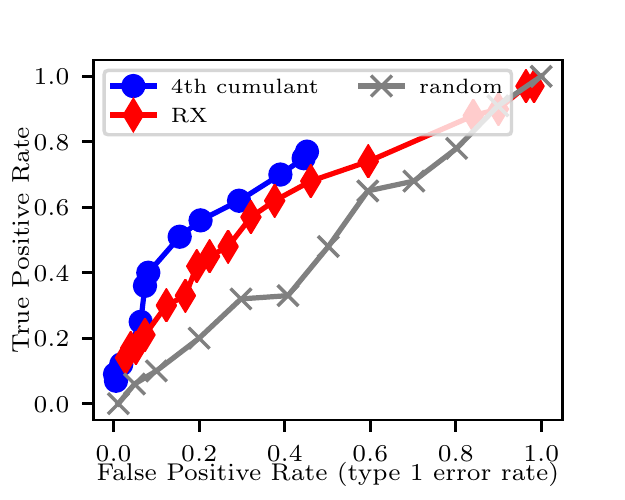}}
	\subfigure[$\nu_c = 
	8$\label{fig::nu8}]{\includegraphics[width=0.32\textwidth]{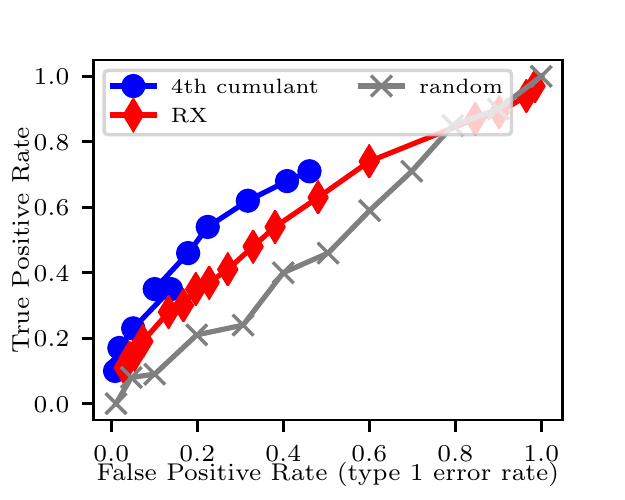}}
	\subfigure[$\nu_c = 
	10$\label{fig::nu10}]{\includegraphics[width=0.32\textwidth]{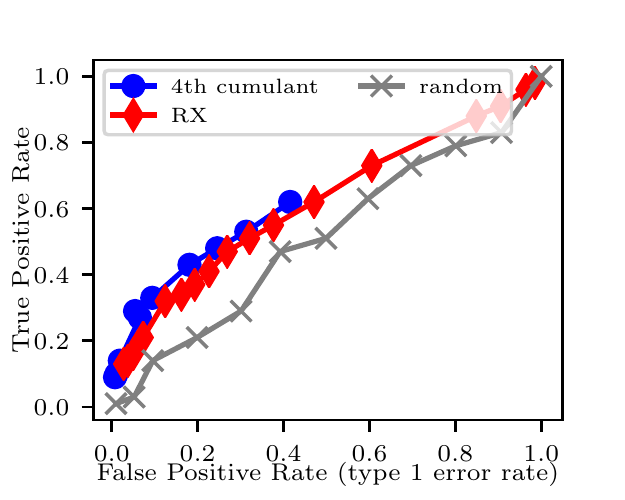}}
	\caption{Exemplary detections of artificial outliers. One can see that the 
	$4$th cumulant outperforms both the RX and the random choice. Although the 
	$4$th cumulant method does not fully cover the detection range, the 
	performance is similar to the mentioned further example of the financial 
	crisis 
	detection. For the cumulant method the number of specific directions 
	parameter was chosen to be $r 
	= 3$.}
	\label{fig::detect}
\end{figure}

The example realisations of the detection are presented in 
Figure~\ref{fig::detect}, for more examples run tests in 
\texttt{./test/outliers\_detect/} of~\cite{cumf}. The results 
are presented in the form of the ROC 
(Receiver Operating Characteristic) curve. The True Positive Rate is the rate 
of truly detected outliers out of all outliers. The False Positive Rate is the 
rate of the type one error (the number of the falsely detected data over the 
number of the ordinary data).

The $4$th cumulant method outperforms the RX and the random choice. It does not 
fully cover the detection range, but it can assure the $60-80\%$ True Positive 
Rate for the low False Positive Rate.
 In the author's opinion, the advantage over the RX detector comes from the
 limitations concerning the use of the Mahalanobis distance in the RX detector 
 for the non-Gaussian distributed data.

\section{Financial data analysis}\label{sub::fint}

\begin{figure}
	\centering
	\subfigure[The DJI index, the red line marks the start of the 
	crisis at January 25, 
	2018.\label{fig::djindex}]{\includegraphics[width=0.49\textwidth]{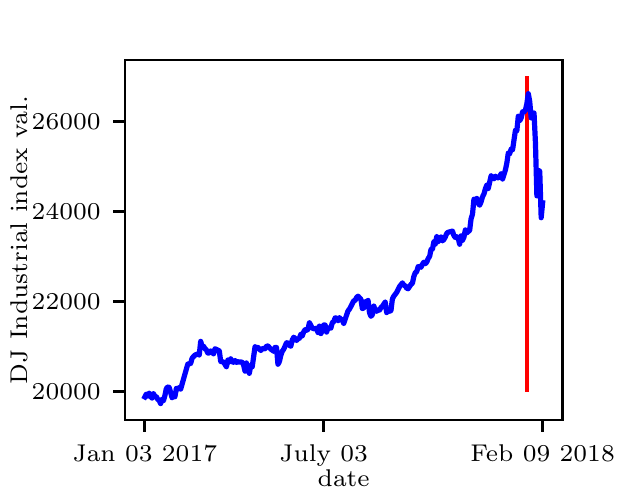}}
	\subfigure[The crisis detection 
	performance, for the cumulant method used $r 
	= 3$ specific 
	directions.\label{fig::djroc}]{\includegraphics[width=0.49\textwidth]{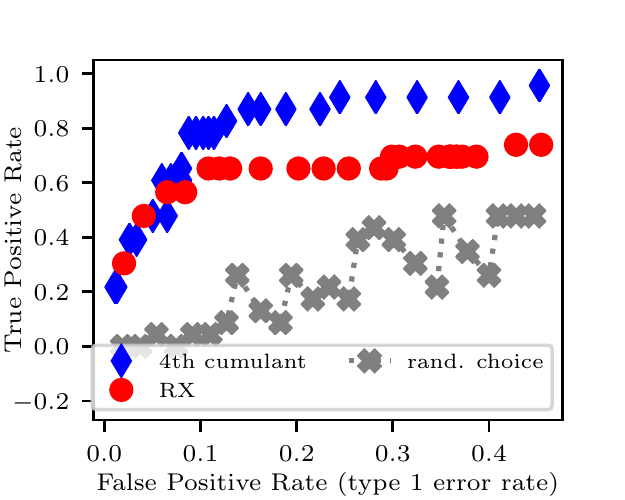}}
	\caption{The crisis detection on New York Stock Exchange. The red vertical 
	line splits data between the ordinary data and the crisis data.}
	\label{fig::dj}
\end{figure}

\begin{figure}
	\centering
	\subfigure[The WIG20 index, the red line marks the start of the 
	crisis at July 28, 
	2011.\label{fig::wigindex}]{\includegraphics[width=0.49\textwidth]{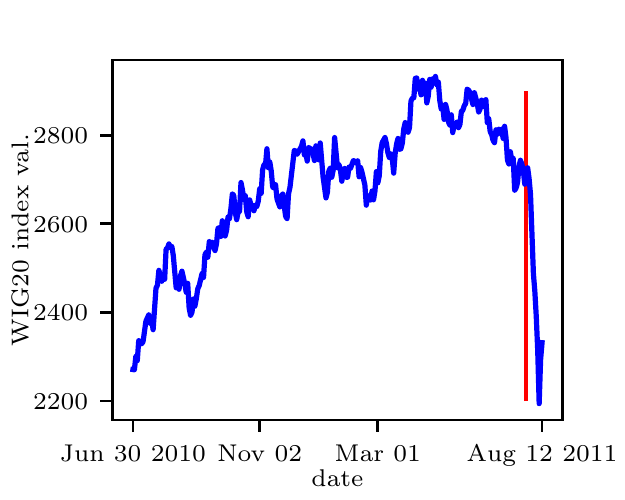}}
	\subfigure[The crisis detection 
	performance, for the cumulant method used $r 
	= 3$ specific 
	directions.\label{fig::wigroc}]{\includegraphics[width=0.49\textwidth]{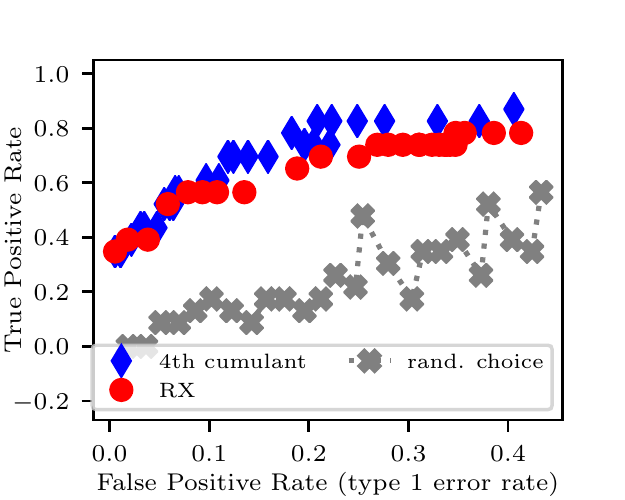}}
	\caption{The crisis detection on Warsaw Stock Exchange. The red vertical 
		line splits data between the ordinary data and the crisis data.}
	\label{fig::wig}
\end{figure}

In this section, the application of the outlier detection on the real life 
financial data  is discussed. The goal of the analysis concerns the crisis 
detection and the crisis prediction in the equity 
markets. For the data representation, individual assets correspond to marginals 
and their log 
increases correspond to the realisations. The assets are share prices of 
important 
companies traded on two distinct stock exchanges: the large 
New York Stock Exchange and the medium size Warsaw Stock Exchange. The reason 
for such selection of data is that one may expect different dynamics of the 
large well-developed market in comparison to the dynamics of the medium-size, 
not fully 
developed market.
In both cases, the analysis concerns the data from approximately $270$ trading 
days, that covers the whole trading year. In both cases, the 
author selected the period when the increase of the 
index value 
was followed by the crisis, i.e.: for the New York Stock 
Exchange January 2017 - February 2018, see Figure~\ref{fig::djindex}; for the 
Warsaw 
Stock Exchange June 2010 - August 2011, see Figure~\ref{fig::wigindex}. To 
increase the number of data for statistics, two 
records per day are 
taken: the opening price and the closing price.

The input data are the log increments of the share prices of the companies 
included in 
the relevant stock 
exchange index. In the case of the New York Stock Exchange, the Dow Jones 
Industrial (DJI) index containing $30$ major companies is used.  
Due to the data availability, the
analysis is limited to the $29$ following companies: AAPL, BA, CSCO, DIS, GS, 
IBM, JNJ, KO, MMM, MSFT,  
PFE, TRV, UTX, VZ, WMT, AXP, CAT, CVX, DWDP, HD, INTC, JPM, MCD, MRK, NKE,   
PG, UNH, V, WBA. In the case of the Warsaw Stock Exchange, the WIG20 
index containing $20$ major companies is used. Analogically, due to the data 
availability, the
analysis is limited to the $19$ following companies: ASSECOPOL, CEZ, GTC, 
KERNEL, 
LOTOS, ORANGEPL, 
PEKAO, PGNIG, PKOBP, TAURONPE, BOGDANKA, GETIN, HANDLOWY, KGHM, 
MBANK, PBG, PGE, PKNORLEN, PZU.

For the crisis detection, scenario the 
observation window includes mostly ordinary trading, but ends on the 
crisis data, see Figure~\ref{fig::djindex} and Figure~\ref{fig::wigindex}. The 
size in the observation window covers the whole trading 
year, hence the seasonal fluctuations of 
the ordinary data are accounted for. The crisis data were selected 
manually, looking for sharp drops of the index values. 
Ordinary and crisis data are separated by the red vertical lines in 
Figure~\ref{fig::djindex} and 
Figure~\ref{fig::wigindex}.

The analysis of the crisis detection performance is presented in 
Figure~\ref{fig::djroc} and Figure~\ref{fig::wigroc}, where the True 
Positive Rate of the detection is plotted against the False Positive Rate being 
the rate of the type one error. In most cases the $4$th 
cumulant algorithm outperforms the RX detector. Hence, both for WIG20 and DJI 
data, the 
higher-order cumulant (of order $4$ in this case) can play the meaningful role 
in the crisis detection. Given the False Positive Rate a bit higher than $0.1$ 
the True Positive Rate of the $4$th cumulant algorithm is large and the crisis 
can be detected with high probability after recording one piece of the crisis 
data. This region corresponds to the detectors sensitively parameter roughly 
fulfilling $\beta \leq 
2.5$, see Algorithm~\ref{alg::det}. However, as the $\beta$ parameter drops,  
the False Positive rate rises.

Furthermore, referring to Figure~\ref{fig::wig}, the performance of the crisis 
detection on the Warsaw Stock Exchange 
appears to be similar to the detection of outliers artificially sampled by 
means 
of the t-Student copula, see Figure~\ref{fig::detect}. Hence, the t-Student 
copula appears to be a good 
model of the crisis data of the medium size stock exchange. 
Referring to the New York Stock Exchange case, the performance of the $4$th 
cumulant is even better. This may suggest higher cross-correlation of extreme 
events for the examined crisis on the New York 
Stock Exchange. However, since the $4$th cumulant measures the 
information tied to the t-Student copula (see Section~\ref{s::prel}) the 
t-Student copula (or perhaps its generalisation) should be considered as the 
crisis data model here as well.

\begin{figure}
	\centering
	\subfigure[The WIG20 index, the green line marks the start of the period 
	directly preceeding the crisis (July 11, 2011), the red line marks the 
	start 
	of the 
	crisis (July 28, 
	2011).\label{fig::wigindexpred}]{\includegraphics[width=0.49\textwidth]{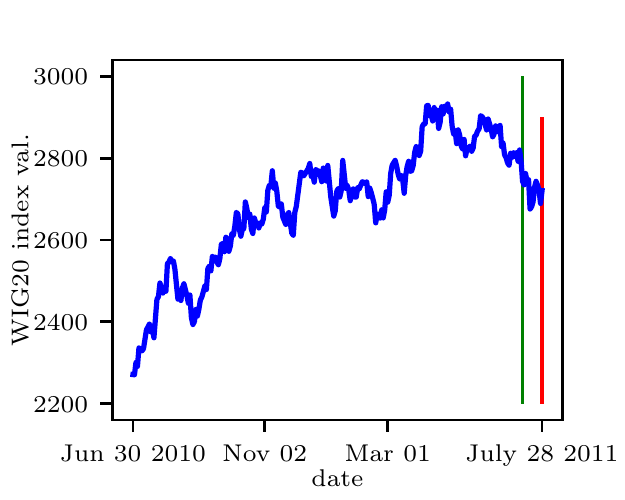}}
	\subfigure[The crisis prediction, $r 
	= 
	3$.\label{fig::wigrocpred}]{\includegraphics[width=0.49\textwidth]{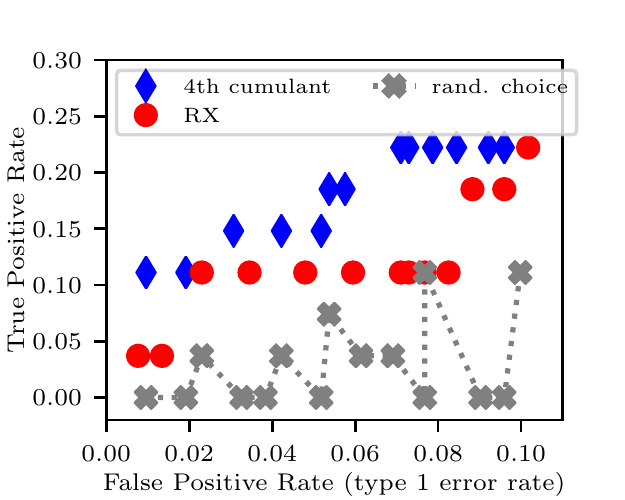}}
	\caption{The crisis prediction on Warsaw Stock Exchange. The range with the 
		low False Positive Rate was chosen, since we have the sequence of 
		pre-crisis data to make a proper detection. The advantage of the $4$th 
		cumulant is for False Positive Rate smaller than the $0.1$, that 
		correspond to the 
		parameter $\beta \geq 2.75$.}
	\label{fig::wigpred}
\end{figure}

Since the Warsaw Stock Exchange example fits to the 
theoretical model, it will be chosen for the crisis prediction scenario. In 
this scenario, the green line separates the ordinary data from the outliers 
(crisis predecessors), see 
Figure~\ref{fig::wigpred}.  Hence, there are several pre-crisis events 
available before the crisis occurs, one should concentrate rather on the low 
False Positive Rate. In this region, the $4$th cumulant outperforms the RX 
detector, see Figure~\ref{fig::wigrocpred}.


\section{Conclusions}

Given the analysis of the New York Stock Exchange and the Warsaw Stock 
Exchange, one can conclude that the $4$th order multivariate cumulants are 
applicable in the crisis detection scenario both on large and medium sized 
stock 
exchanges. These cumulants measure the
cross-correlation of the extreme events, what appears during the crisis. Such 
extreme events can be modelled by the t-Student copula that is informatively 
tied to $4$th order multivariate cumulants.
Due to the analogy in outliers detection scheme, the t-Student copula is best 
applicable to model the crisis on the medium sized stock exchanges such as the 
Warsaw Stock Exchange. 

Interesting aspect concerns the feasibility of the $4$th order multivariate 
cumulants in the crisis 
prediction. Here, positive results for the Warsaw Stock Exchange rise the 
suggestion that in 
analogy to the auto-correlated extreme 
events, the cross-correlated extreme events may appear before the crisis. Since
the cumulant based outliers detector does not look into the 
auto-correlation of financial data it may support the Hurst Exponent based 
methods looking into auto-correlation of these data for crisis prediction.

\subsubsection*{Acknowledgments}
The research was partially financed by the National Science Centre, Poland --
project number 2014/15/B/ST6/05204. The author would like to thank Jaros{\l}aw 
Adam Miszczak for revising the manuscript and discussion and Adam Glos for an 
implementation assistance.

\end{document}